\documentclass[appendixfloats,iop]{emulateapj}
\usepackage{longtable}
\usepackage{xcolor}
\begin{document}

\title{A Break in Spiral Galaxy Scaling Relations at the Upper Limit of Galaxy Mass}

\author{Patrick M. Ogle$^1$, Thomas Jarrett$^2$, Lauranne Lanz$^3$, Michelle Cluver$^{4,5}$, Katherine Alatalo$^1$,  Philip N. Appleton$^6$, Joseph M. Mazzarella$^6$}

\affil{$^1$Space Telescope Science Institute, Baltimore, Maryland}

\affil{$^2$University of Cape Town, South Africa}

\affil{$^3$The College of New Jersey, Ewing, New Jersey}

\affil{$^4$ Swinburne University of Technology, Melbourne, Australia}

\affil{$^5$  Department of Physics and Astronomy, University of the Western Cape, South Africa}

\affil{$^6$ IPAC, California Institute of Technology, Pasadena, CA}

\email{pogle@stsci.edu}

\shorttitle{Cosmic Mass Limit for Galaxies} 
\shortauthors{Ogle et al.}

\begin{abstract}
Super spirals are the most massive star-forming disk galaxies in the universe \citep{oln16, o19}.  We measured rotation curves for 23 massive spirals\footnote{Based on observations made with the Southern African Large Telescope (SALT)} and find a wide range of fast rotation speeds (240-570 km s$^{-1}$), indicating enclosed dynamical masses of  $0.6- 4 \times 10^{12} M_\odot$.  Super spirals with mass in stars $\log M_\mathrm{stars} /  M_\odot  > 11.5$ break from the baryonic Tully-Fisher relation (BTFR) established for lower mass galaxies. The BTFR power-law index  breaks from $3.75 \pm 0.11$ to $0.25 \pm 0.41$ above a rotation speed of $\sim 340$ km s$^{-1}$.  Super spirals also have very high specific angular momenta that break from the \cite{f83} relation. These results indicate that super spirals are undermassive for their dark matter halos, limited to a mass in stars of $\log M_\mathrm{stars} / M_\odot < 11.8$. Most giant elliptical galaxies also obey this fundamental limit, which corresponds to a critical dark halo mass of $\log M_\mathrm{halo}  / M_\odot \simeq 12.7$. Once a halo reaches this mass, its gas can no longer cool and collapse in a dynamical time.  Super spirals survive today in halos as massive as $\log M_\mathrm{halo} / M_\odot \simeq 13.6$, continuing to form stars from the cold baryons they captured before their halos reached critical mass.  The observed high-mass break in the BTFR is inconsistent with the Modified Newtonian Dynamics (MOND) theory \citep{bm84}.  

\end{abstract}

\section{Introduction}

Super spiral galaxies are extreme by many measures, with r-band luminosities of $L=8-14 L^*$, stellar masses of $M_\mathrm{stars} = 2-6\times 10^{11} M_\odot$, and giant isophotal diameters of $D_{25} = 55-134$ kpc  \citep{oln16,o19}.  They represent a very rare population of massive disk galaxies in which star formation has not quenched.  As such, they provide a unique opportunity to extend studies of galaxy scaling laws into an entirely new regime. 

The discovery of flat, high-velocity rotation curves firmly established the presence of dark matter in galaxies \citep{b78,r78}.  Dark matter halos \citep{wr78,nfw97,gnc08} are fundamental to galaxy formation, forming the 
scaffolding for gas accretion and star formation.    Though the composition of dark matter remains unknown, it is a crucial component of  $\Lambda$-Cold Dark Matter ($\Lambda$CDM)
cosmology,  describing the expansion history of the universe.  
The rotational angular momenta of galaxies may be imparted by torques on their primordial dark matter halos by the surrounding irregular matter distribution prior to their collapse \citep{f79,fe80}.  When the baryons cool and collapse, they spin up,  retaining most of their original angular momentum. The specific angular momentum of galaxies generally increases following the \cite{f83} relation $j* \sim M_{stars}^{0.6}$.

 The high rotation speeds of galaxies have alternatively been attributed to a breakdown in Newtonian dynamics in the regime of low gravitational acceleration \citep{m83, l17}.  In particular,  the Modified Newtonian Dynamics theory (MOND) suggests a specific form for the gravitational potential that leads to flat rotation curves and obviates the need for dark matter \citep{bm84}.  

The Tully-Fisher relation \citep[TFR:][]{tf77} between galaxy optical ({$B$-band) luminosity and H {\sc I} line width has played an important role in galaxy evolution studies and mapping galaxy peculiar velocities in the local universe \citep[e.g.,][]{wcf97,hgc99,fzc99,smh07,tcs16}.  Substituting I-band or mid-infrared (MIR) for $B$-band photometry reduces the scatter in the TFR because of reduced extinction and smaller scatter in mass-to-light ratio \citep[e.g.,][]{gh94,tc12, lms13,sct13,nst14,zcm14,l16}. The resultant infrared Tully-Fisher relation (ITFR) relates mass in stars $M_\mathrm{stars}$ to the rotation velocity and gravitational potential of the baryonic plus dark dynamical mass within radius $r$: $M_\mathrm{dyn}(r) \sim r v_\mathrm{rot}^2$.  

For spiral galaxies with $M_\mathrm{stars} = 10^{10}-10^{11} M_\odot$, the ITFR has a power-law index of $3.75 \pm 0.11$ \citep{l16}.  This is greater than the index of 3.0 that would be predicted under the assumption of constant stellar mass fraction \citep{m12}, indicating that star-formation efficiency increases with $M_\mathrm{dyn}$ for galaxies in this mass range.  Massive spiral galaxies with $M_\mathrm{stars} \sim 10^{11} M_\odot$ may be the most efficient at converting gas into stars, with low gas fractions and high stellar mass fractions that approach the cosmic baryon fraction of 0.167 \citep{pfm19,pmff19, kdr09}.  Adding in the neutral gas masses of spirals gives a tighter relation that removes the low-mass break in the TFR  for dwarf galaxies \citep{m00, m12, l16}.  This baryonic Tully-Fisher relation (BTFR) demonstrates a strong connection between the cold baryonic (stars $+$ cold atomic and molecular gas) and dark matter content of spiral galaxies. The slope and scatter of the BTFR  depend in detail on the prescription used to estimate the mass-to-light ratio of stars and which galaxy samples are selected \citep{ms15,sg16,pvp18}. Evidence for a  flatter ITFR  slope is found for high-redshift  galaxies \citep{ch17}.  However, MOND predicts a BTFR slope of exactly 4.0, and any deviation from this is at odds with that theory.

The shape of the TFR should reflect that of the stellar-mass/halo-mass (SMHM) relation, constructed by matching galaxies drawn from the observed luminosity function to simulated dark halos and subhalos \citep{kbw04, hbc10, mnw13, bwc13}.  The SMHM relation has a characteristic break at $\log M_\mathrm{stars} / M_\odot \simeq 10.5$, corresponding to the observed break in the \cite{s76}  luminosity function at $L^*$. Since the dark matter halo mass function is scale-invariant, this break in star formation efficiency must reflect the baryonic physics of galaxy formation and evolution.  It is commonly attributed to a transition from stellar feedback to AGN feedback dominance \citep[e.g.,][]{db06,csw06,s15,shh19}.   However, the BTFR does not show a break at the same scale \citep{tg11,d12}, pointing to a different SMHM relation for spirals and ellipticals \citep{pmff19}. 

Previous studies of the BTFR have been limited to galaxies with stellar masses $<2\times 10^{11} M_\odot$, because galaxies with higher masses are quite rare and because it is difficult to detect
H {\sc i} at $z>0.1$. The extreme stellar masses and sizes of super spirals allow us to probe spiral disk dynamics and massive galaxy dark matter halos at radii of up to $\sim 50$ kpc. We use optical 
long-slit spectroscopy of the H$\alpha$ line to measure the rotation curves of 23 massive spirals and place them on the BTFR.

We assume a $\Lambda$CDM cosmology with $H_0=70$ km s$^{-1}$ Mpc$^{-1}$, $\Lambda=0.7$ and $\Omega=0.3$  to derive all distances, linear sizes, and luminosities.

\section{Sample and Observations}

We selected our rotation curve sample (Table 1) from parent samples selected by $r$-band or  $K_s$-band luminosity.  First, we selected galaxies with inclination $i>39\arcdeg$ from the OGC sample of super spirals with $z<0.3$ and Sloan Digital Sky Survey (SDSS) $r$-band luminosity $>8 L*$ \citep{o19}.  Next, because extinction in the disk limits the number of high-inclination galaxies in the OGC, we created a new  sample of IR-selected massive spirals drawn from the set of 2 Micron All-Sky Survey Extended Source Catalog (2MASX) galaxies with SDSS-measured redshifts,  $i>39\arcdeg$,  $K_s$-band luminosity $ L(K_s) >  2\times 10^{11} L_\odot (K_s)$, and $r$-band isophotal diameter $D_{25} > 50$ kpc.  The $K_s$-band luminosity and $D_{25}$ criteria  were designed to yield a sample that overlaps \cite{o19} super spirals, which have $M_\mathrm{stars}>2\times 10^{11} M_\odot$ and $D_{25}>55$ kpc.  

We observed 3 massive spirals with the Double Spectrograph (DBSP) on the Hale Telescope and 20 with the Robert Stobie Spectrograph \citep[RSS:][]{bnk03,knb03} on the Southern African Large Telescope \citep[SALT:][]{bsm06}\footnote{SALT programs 2018-2-SCI-027, 2019-1-SCI-028, PI: T. Jarrett}.   We placed a $1\arcsec$ wide, long slit along the $r$-band major axis of each galaxy.  We used the DBSP red 1200 line mm$^{-1}$ grating, 
yielding a dispersion of 0.300 $\AA$ pixel$^{-1}$  and a resolving power of 6000. The spectral resolution is 50 km s$^{-1}$ at H$\alpha$ and the plate scale is $0\farcs293$ pixel$^{-1}$.  At SALT, we used the RSS p1800 line 
mm$^{-1}$ holographic grating together with the pc04600 order blocking filter. With 2x2 pixel  on-camera binning, this  grating gives a resolving power of $4200-5300$.  The spectral resolution is 71 - 57 km s$^{-1}$  at H$\alpha$ and the plate scale is  $0\farcs254$ pixel$^{-1}$.

Exposures were median-combined to remove cosmic ray tracks, yielding exposure times of 30-70 minutes. Spectra were rectified and
wavelength-calibrated using night-sky lines.  We  subtracted sky foreground emission using regions above and below the  spectra. Galaxy continuum 
profiles were measured from adjacent spectral continuum regions, scaled and subtracted. The resulting 2D spectra were median-filtered with a 3x3 pixel kernel to improve S/N.
We measured the rotation curve (Fig. 1) from the weighted centroid wavelength of the H$\alpha$ emission line, up-weighting high-surface brightness regions to mitigate the effect of dilution
from regions at lower projected velocity inside the slit. This diluting emission can be seen as fainter emission at lower velocity (Fig. 1b). The zero point of the rotation curve was set to minimize the asymmetry between the approaching and receding sides.  The rotation curve was then sampled at intervals of $1\farcs0-1\farcs5$, to match the seeing conditions and the maximum rotation speed $v_\mathrm{max}$ measured from the sampled rotation curve. The standard deviation of the difference between the rotation curve and its spline interpolation gives the uncertainty in the rotation speed at the sampled points.    The smaller linear size of the H$\alpha$ disks compared to the H I disks of spiral galaxies does not produce any significant difference between their optical and H {\sc i} velocity widths \citep{kff02}, allowing a direct comparison.

\section{Rotation Speed and Dark Matter Content}


\begin{figure}[ht]
  \includegraphics[trim=0.0cm 4.5cm 1.0cm 4.5cm, clip, width=\linewidth]{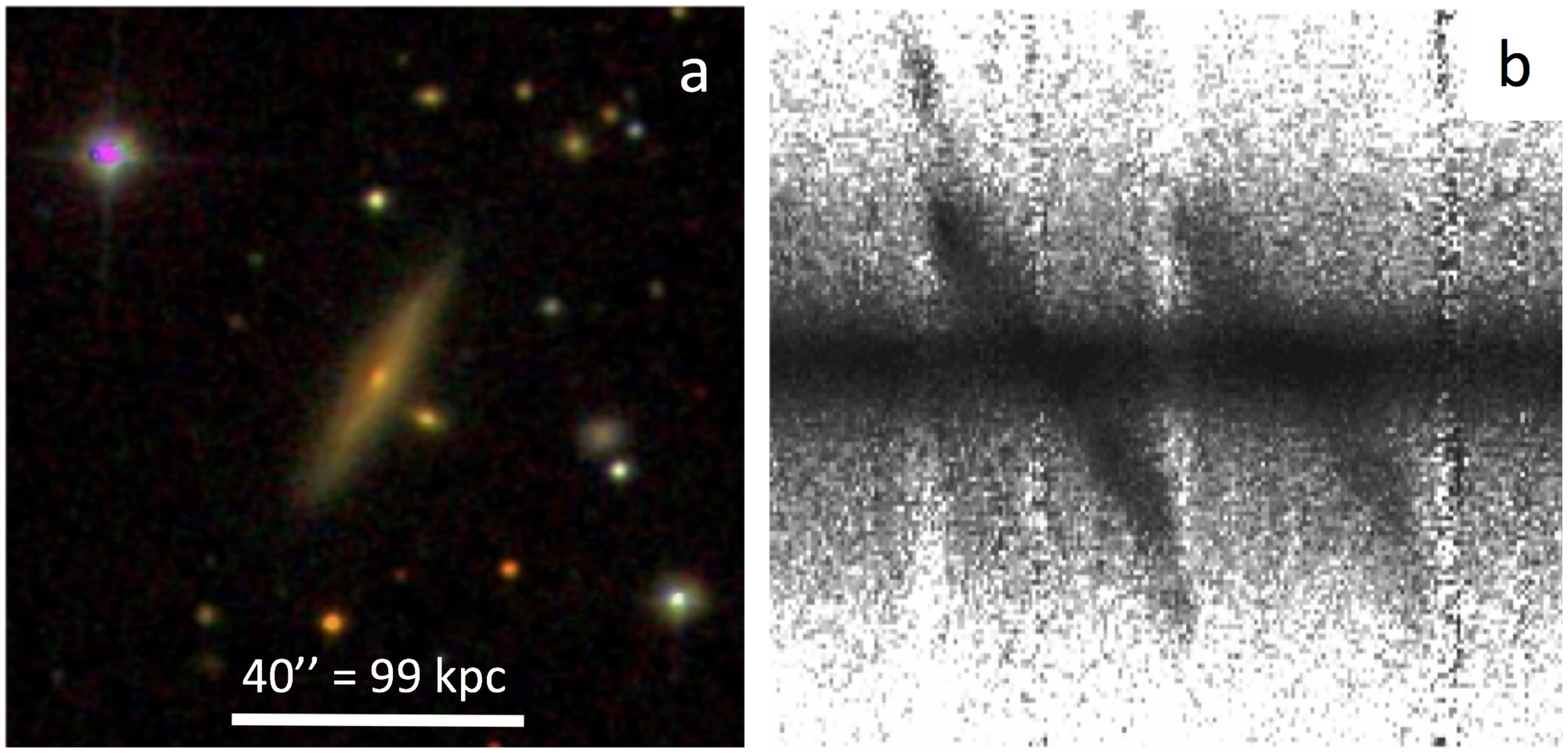}
 \includegraphics[trim=4.0cm 1.5cm 5.8cm 2.0cm, clip, width=\linewidth]{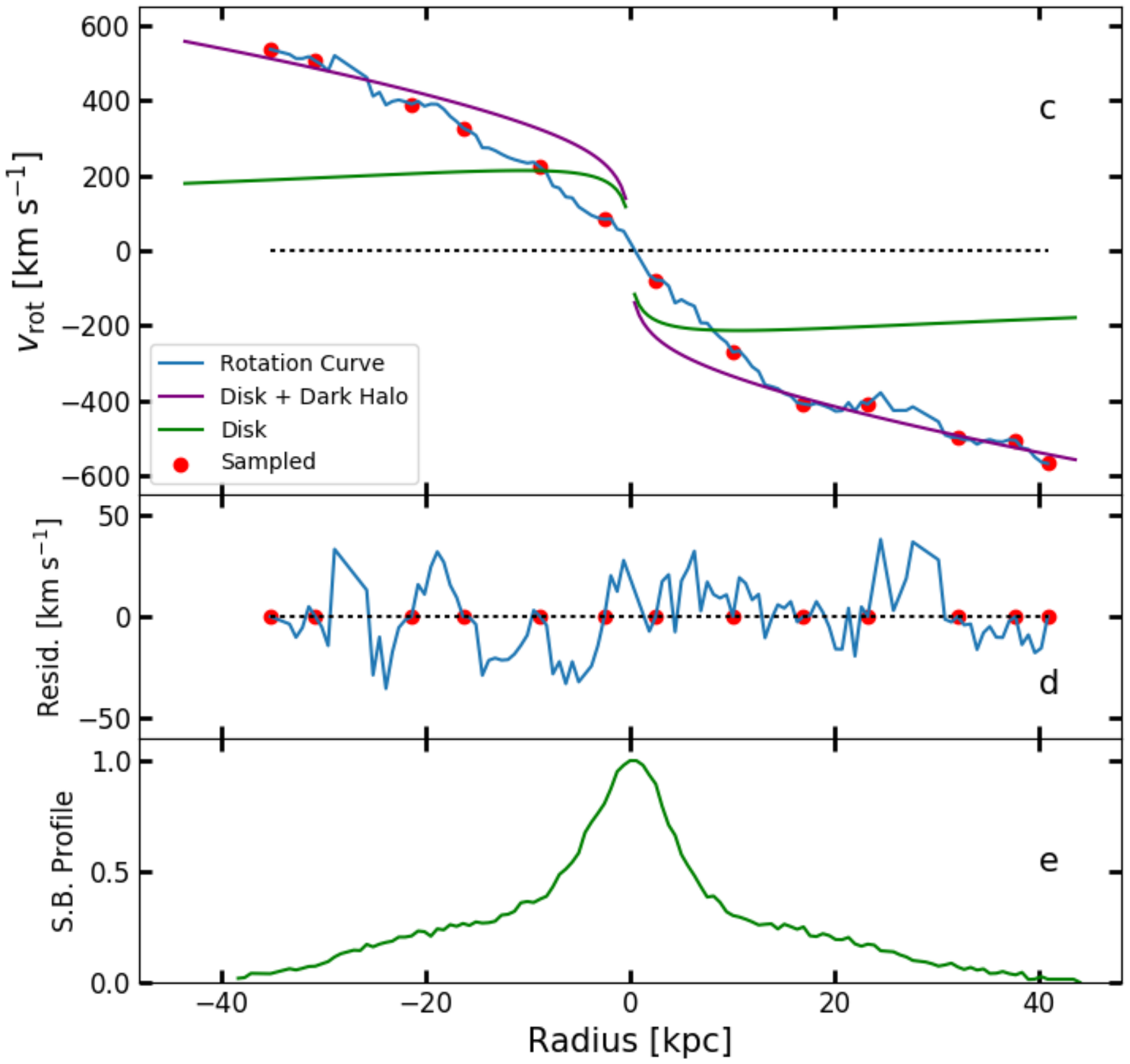}
   \figcaption{Fastest rotating super spiral 2MFGC 12344. a) SDSS $g, r, i-$ band image. b) SALT RSS 2D spectrum of H$\alpha$ and [N {\sc ii}] $\lambda 6585$.  c) Rotation curve 
   (solid blue line, red points) with  $v_\mathrm{max} = 568 \pm 16$ km s$^{-1}$ at  $r = 41$ kpc.   
    The model (purple) includes a disk of stars and gas (green curve) following the observed galaxy surface brightness profile plus NFW dark-matter halo.        
    d)  Residuals to cubic spline fit, with standard deviation 16 km s$^{-1}$.  e) Galaxy continuum profile along the slit.}                     
    \label{1}
\end{figure}
  
We find de-projected super spiral maximum rotation speeds of $v_\mathrm{max} = $ 243-568 km s$^{-1}$ at radii of $r = 14-54$ kpc (Table 1). The rotation curves of most super spirals follow the typical pattern of rising from the 
galaxy center, then flattening at large radii. In two cases (OGC 1304 and OGC 0586), deviations from regular rotation are seen, indicating that the disks may be warped at their outer edges. We conservatively discard these edge points before measuring $v_\mathrm{max}$.  The rotation curves of the two largest, most massive galaxies (OGC 0139 and 2MFGC 12344) continue to rise at the outer edge.  This may lead to an underestimate of the maximum circular velocity.  The galaxy OGC 0139 also has a high uncertainty of 90 km s$^{-1}$ in $v_\mathrm{max}$ because of the large velocity dispersion in two blobs at either edge of its rotation curve.

We separately integrate the gravitational potential from stars and gas  (both assumed to lie in a thin disk) and a spherical dark matter halo following a Navarro, Frenk, \& White (NFW) density profile \citep{nfw97} and compare the resulting model rotation curves to the observed rotation curves (Fig. 1(c)). The contribution of dark matter to the rotation curve of  the fastest rotator, 2MFGC 12344, exceeds the contribution from stars at radii $r>10$ kpc and continues to rise out to $41$ kpc.  The model over-predicts the rotation speed at  $r<15$ kpc, indicating that there is a deficit of rotational support, with gas following non-circular orbits in this inner region of the galaxy, perhaps indicating the presence of a stellar bar  \citep[e.g.,][]{kdw06}.  We modeled the rotation curves of all galaxies in our sample using this same method to estimate the mass of dark matter ($M_\mathrm{dark}$) within the galaxy (Table 1).  We find a large range of $M_\mathrm{dark}=0.25-3.0\times10^{12} M_\odot$ inside the region probed by the rotation curves, corresponding to the large range in maximum rotation speed.

\begin{figure*}
  \includegraphics[trim=0.0cm 4.0cm 0.5cm 4.2cm, clip, width=\linewidth]{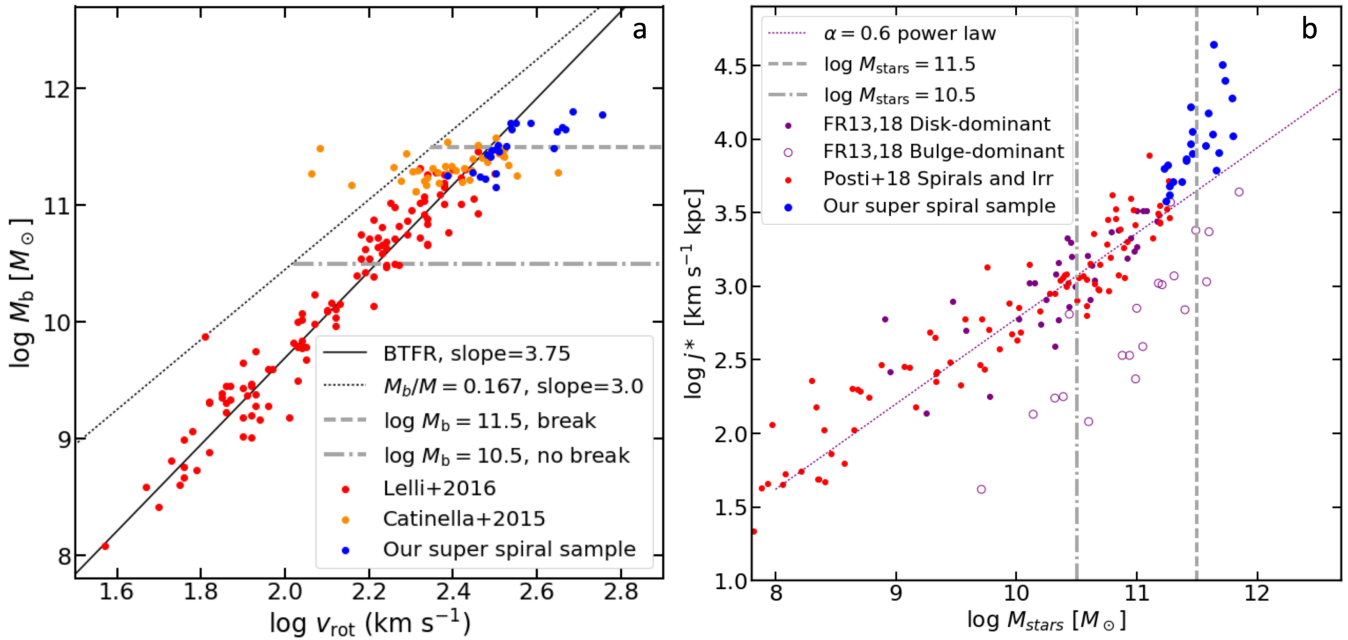}
   \figcaption{Baryonic Tully-Fisher relation (BTFR) and \cite{f83} relation.  A break in both relations is found at a critical stellar mass of $\log M_\mathrm{stars}/ M_\odot = 11.5$ (dashed lines).  This is a factor of
   10 greater than the characteristic mass of  $\log M_\mathrm{stars}/ M_\odot = 10.5$  at the break in the galaxy SMHM relation.
    a) BTFR. Masses in stars for the super spiral and comparison samples are estimated using custom {\it WISE} $W1$-band photometry,
    assuming $M/L=0.6$. The photometric uncertainty is smaller than the size of the plot symbols  (0.01-0.02 dex). Gas masses for the comparison samples are estimated as 
    $M_\mathrm{gas} = 1.33 \times M_\mathrm{HI}$  \citep{l16,c15}, while gas masses for our sample are estimated using the \cite{k98} Schmidt law, with uncertainties $<0.05$ dex (see main text).  
    The observed BTFR (data points) is  compared to the \cite{l16} power-law fit (solid line) and the $v_\mathrm{rot}^3$ power-law for baryon fraction equal to the cosmic mean value (dotted line).  
    b) \cite{f83} relation between galaxy specific angular momentum and mass in stars. The specific angular momenta of our sample galaxies are estimated by $j_* = 2 R_d v_{max}$.  We compare to disk-dominant
        spirals with bulge-to-disk mass ratios $\beta_* < 0.15$ and bulge-dominant ($\beta_* >0.70$) ellipticals from \cite{fr13, fr18} and spirals and dwarf irregulars from \cite{pfd18}.   Super spirals have exceedingly 
        high specific angular momenta compared to lower mass spirals and deviate from the Fall relation \citep[purple dotted line;][]{fr18}.  The relation for elliptical galaxies is steeper and offset to lower $j_*$. 
         \label{2}}
\end{figure*}

\begin{figure}
  \includegraphics[trim=3.0cm 0.0cm 3.0cm 0.0cm, clip, width=\linewidth]{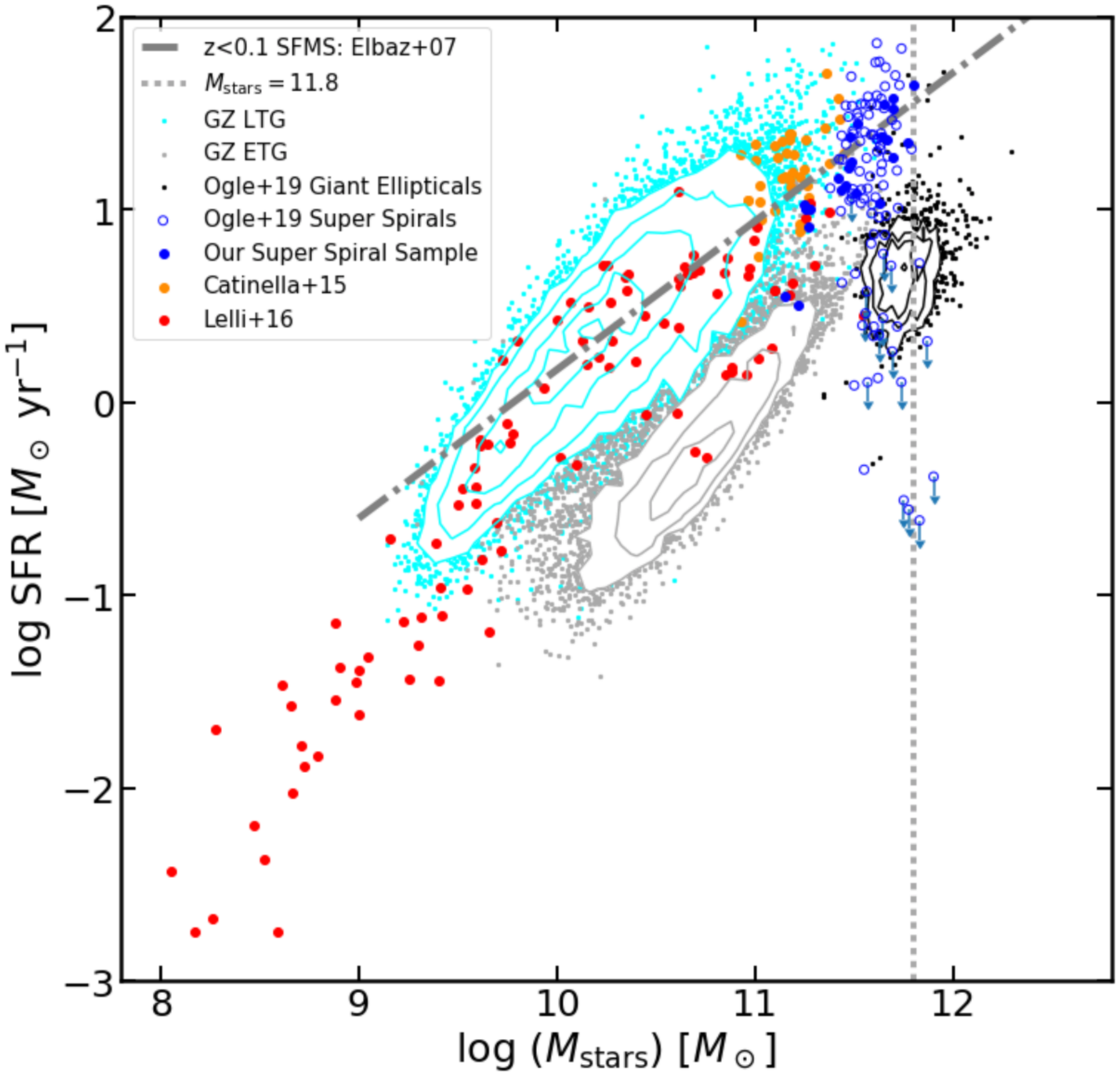}
   \figcaption{Star-forming main sequence (SFMS) \citep[adapted from][]{o19}, with our rotation curve sample, BTFR comparison samples, and Galaxy Zoo (GZ) early type and late type galaxies over-plotted. 
    The observed cosmic mass limit for spiral galaxies at $\log M_\mathrm{stars} = 11.8$ is indicated by the vertical dotted line.  Super spirals fall along an extrapolation of the \cite{e07} relation.  
     Most giant ellipticals and lenticulars in the \cite{o19} sample respect the cosmic mass limit for spiral galaxies. The ones that do not may be the product of major mergers.
        \label{3}}
\end{figure}

\section{Baryonic Mass}

Estimates of total mass in stars from galaxy luminosity depend on filter bandpass and assumed mass to light ratio, which have a direct effect on the overall normalization of the TFR \citep{ms15}.
We use two methods to estimate $M_\mathrm{stars}$ and compare their scatter relative to one another and the resulting scatter in the TFR.  First, we estimate  $M_\mathrm{stars}$ from our custom 
WISE $W1$-band (3.4 $\mu$m) photometry, assuming $M_\mathrm{stars}/L_{W1} = 0.6$.  Our measurements are in good agreement with $M_\mathrm{stars}$ estimated by \cite{l16} for their sample from {\it Spitzer} 
IRAC [3.6]-band luminosity and $M_\mathrm{stars}/L_{3.6} = 0.5$, with scatter driven by photometric uncertainties of $2-5\%$. Next, we estimate $M_\mathrm{stars}$  from WISE $W1$-band luminosity and $M_\mathrm{stars}/L_{W1} $ estimated from $W1-W2$ color, using the prescription of \cite{cjh14}. This empirical relation was derived by comparing $W1$-band luminosity to  stellar mass estimated via stellar-population synthesis models \citep{thb11}.  This estimate is systematically offset from the constant $M/L$ estimate, yielding lower $M_\mathrm{stars}$ values  and a significant difference in slope for the TFR in the mass range $\log M_\mathrm{stars} / M_\odot = 10.0-11.0$.  Because it results in lower scatter for the ITFR over the full mass range probed, we use our WISE $W1$-band  $M_\mathrm{stars}$ estimates with constant $M_\mathrm{stars}/L_{W1} = 0.6$ for all galaxies in our sample and comparison samples. Based on $W1 - W2$ color, only 2MFGC 10372 has significant AGN contamination of its $W2$- and $W3$-band flux measurements. This has no significant impact on our $W1$-based estimates of
$M_\mathrm{stars}$ and $M_b$, but we regard our $W3$-based SFR and $M_\mathrm{gas}$ estimates for this galaxy as upper limits.

Our efforts to measure the H {\sc I} masses of super spirals at Arecibo and with the Green Bank Telescope (GBT) have so far been thwarted by strong radio-frequency interference (RFI) at
their redshifted H {\sc I} frequencies.  Instead, we estimate the mass in cold gas $M_\mathrm{gas}$ via the \cite{k98} Schmidt (KS) law, which relates the SFR and cold gas mass (H {\sc I} $+$  H$_2$) surface densities. We estimate SFR surface density from WISE $W3$-band luminosity, using the prescription of \cite{cjd17}, and the $r$-band isophotal diameter $D_{25}$.  This results in gas masses of  $\log M_\mathrm{gas} = 9.9- 10.9$ (Table 1) and gas fractions of $M_\mathrm{gas}/M_\mathrm{b} = 0.05-0.14$.  The relatively low cold gas fractions are typical for massive spiral galaxies \citep{l16}.  Using the same method for the gas-rich \cite{c15} sample, we find that they fall on average 0.36 dex to the right (and below) the KS relation.  We adopt an uncertainty for our KS-derived $M_\mathrm{gas}$ values equal to the observed scatter of 0.3 dex in the KS relation \citep{k98}, which translates to an uncertainty of only 0.05 dex in $M_\mathrm{b}$.

\section{Discussion and Conclusions}

\subsection{Baryonic Tully-Fisher and Fall Relations}

The BTFR (Fig. 2(a)) relates the total baryonic mass ($M_\mathrm{b} = M_\mathrm{stars} + M_\mathrm{gas}$) to the dynamical mass at  the radius where the rotation curve becomes flat. It is instructive to compare the observed
BTFR to the expectation for a baryon fraction $f_\mathrm{b}$ equal to the cosmic mean baryon fraction of $f_\mathrm{c} = 0.167$ \citep{kdr09}, which would yield a logarithmic slope of 3.0. Real galaxies fall short of this, with  $ f_\mathrm{b}/f_\mathrm{c} = M_\mathrm{b} /(0.167 M)$ in the range 0.1-0.5, depending on galaxy mass.  Dwarf galaxies with $\log M_\mathrm{stars} = 8.0 $ retain only 10\% of their share of baryons and are the least efficient, converting only 20\%  of that into stars.   Spiral galaxies with $\log M_\mathrm{stars} \sim 11.0$  hold onto 50\% of their cosmic share of baryons and are the most efficient, converting up to 70\% of that into stars.  

We find that super spirals with $v_\mathrm{rot}>340$ km s$^{-1}$ deviate from the established BTFR, with relatively low $M_\mathrm{b}$ for their high rotation velocities (Fig. 2(a)). Including these galaxies, 
the BTFR breaks at a characteristic mass scale of $\log M_\mathrm{b} / M_\odot \simeq 11.5$. Fitting the BTFR for the 10 fastest rotating super spirals gives a power-law slope of $0.25 \pm 0.41$ that is much flatter than the low-mass BTFR slope of  $3.75\pm 0.11$ found by \cite{l16}.  Fitting all 23 massive spirals gives  a power-law slope of $1.64\pm 0.30$.  The slopes for these two fits differ by $8\sigma$ and $7\sigma$, respectively, from the low-mass BTFR slope. 
We emphasize that this high-mass break in the BTFR was not readily apparent before the discovery of extremely massive, fast-spinning super spirals, which are extremely rare \citep{oln16,o19}.

The large departure of super spirals from a power-law BTFR  with slope 4.0 is inconsistent with MOND.  The only way to reconcile our observations with MOND is a large mass of un-observed baryons
inside the radii probed by our rotation curves.  The fastest rotator in our sample (2MFGC 12344) would require $\log M_\mathrm{b} / M_\odot = 12.5$ and $\log M_\mathrm{gas} / M_\odot = 12.4$, factors of  5 and 50 greater than our estimates, respectively, to match the low-mass BTFR.  The observed radial acceleration of $ a = v_\mathrm{max}^2/ r = 0.7-2.5 \times 10^{-10}$ m s$^{-2}$ deviates significantly from the prediction of Newtonian mechanics if there is no dark matter, and is close to the characteristic acceleration scale for MOND \citep[$a_0=1.3\pm0.3 \times 10^{-10}$ m s$^{-2}$, ][]{l17}.  Hence, super spirals are probing a regime where MOND would apply if it were correct. However, the radial accelerations observed in super spiral disks are {\it greater} than the MOND prediction, reflecting their high rotation speeds and deviation to the right of the MOND-predicted BTFR.

\begin{deluxetable*}{llcrrcccccrc}
\tablecaption{Super Spiral Sample}
\tablehead{
\colhead{Name} &\colhead{Alt. Name} & \colhead{$z$\tablenotemark{a}}& \colhead{$D_{25}$\tablenotemark{b} } & \colhead{$R_d$ \tablenotemark{c}} & \colhead{$i$ \tablenotemark{c}} &\colhead{$\log M_\mathrm{dark}$\tablenotemark{d}} &\colhead{$\log M_\mathrm{stars}$\tablenotemark{e}} &\colhead{$\log M_\mathrm{gas}$\tablenotemark{f}} &\colhead{$\log$ SFR\tablenotemark{g}}& \colhead{$v_\mathrm{max}$\tablenotemark{h}}  & \colhead{$r$ (kpc)\tablenotemark{h}}}

\startdata
2MASX J09394584$+$0845033     & \nodata           & 0.13674  & 76   & 11.1 & 62  & 11.9  & 11.45  & 10.7   & 1.44   & 322  ( 9)  & 14  \\  
SDSS J095727.02$+$083501.7     & OGC 0441      &  0.25652  & 88   & 17.0 & 39 & 12.1 & 11.60   & 10.4   & 1.03  & 444 (15) &  31 \\   
2MASX J10222648$+$0911396     & \nodata           &  0.09130  & 92   & 14.5 & 76 & 11.8  & 11.42  &  10.5  & 1.22   & 311 (12)  & 33  \\  
2MASX J10304263$+$0418219     &OGC 0926       &  0.16092  & 70   &   8.7 & 48 & 11.7  & 11.66  & 10.7   & 1.53  & 342 (12) &  30  \\ 
2MASX J11052843$+$0736413    & 2MFGC 08638 &  0.15229  & 144 & 47.0 & 85 & 12.5 & 11.59   & 10.8  & 1.38   & 465 (13) & 54  \\  
2MASX J11232039$+$0018029     & \nodata            &  0.14454 & 104 & 18.9 & 79 & 12.1  & 11.43  &  10.6  & 1.25   & 436 (11)   & 45  \\
2MASX J11483552$+$0325268     & \nodata            &  0.11984 & 88   & 17.3 & 80 & 11.9  & 11.42   & 10.5  & 1.13    & 324 (12) & 31 \\   
2MASX J11535621$+$4923562     & OGC 0586       &  0.16673 & 90   & 15.7 & 63 & 11.5  & 11.64   & 10.8   & 1.58   & 305 (11)  & 19 \\ 
2MASX J12422564$+$0056492     & \nodata            &  0.07936 & 52   &  6.8  & 54 & 11.4  & 11.24  & 10.2   & 1.01   & 279 (10)   & 14 \\ 
2MASX J12592630$-$0146580     & \nodata             &  0.08311 & 67   & 10.4 & 61 & 11.8  &  11.23  & 10.2  & 0.91   &  318 (10)  & 20 \\  
2MASX J13033075$-$0214004     & 2MFGC 10372 & 0.08425  & 71   &  8.3  & 82 & 11.8  & 11.37   & $<10.4$  & $<1.16$    & 308 ( 8) & 23 \\  
SDSS J143447.86$+$020228.6     & OGC 1312       &  0.27991 & 75   & 11.6 & 63 & 11.9   & 11.60  & 10.7   & 1.55  & 344  ( 6)  & 26  \\  
2MASX J15154614$+$0235564    & 2MFGC 12344 & 0.14068  & 120 & 21.9 & 81 & 12.5  & 11.74   & 10.7  & 1.35   & 568 (16) & 41 \\   
2MASX J15404057$-$0009331     & \nodata             &  0.07830 & 71  & 12.2 & 67  & 11.8  & 11.39  &  10.4  & 1.10  & 304 ( 8)  & 30 \\    
2MASX J16014061$+$2718161     & OGC 1304      &  0.16440  & 82  & 11.4 & 59 & 12.3  & 11.63  & 10.6   & 1.37  & 453 (25)  & 29 \\   
2MASX J16184003$+$0034367    & \nodata             &  0.16731 & 95  & 24.6 & 77  & 12.0 & 11.67  & 10.6   & 1.28    & 384 (16) & 40  \\ 
2MASX J16394598$+$4609058     & OGC 0139      & 0.24713  & 134 & 33.0 & 76 & 12.2  & 11.74  &  10.9  & 1.65    & 483 (90) & 31 \\     
2MASX J20541957$-$0055204     & \nodata            &  0.21014 & 84   & 14.7 & 66  & 11.9  & 11.41  & 10.6   & 1.38    & 317 (13) & 37 \\   
2MASX J21362206$+$0056519    & \nodata            & 0.10370  & 75   & 11.8 & 65  & 11.8  & 11.47  & 10.4   & 1.08    & 336 (11) & 29 \\    
2MASX J21384311$-$0052162     & \nodata            & 0.08291  & 60   &  8.5 & 58  &  11.8  & 11.20  &   9.9   & 0.50    & 299 (12) & 15  \\   
2MASX J21431882$-$0820164      & \nodata           &  0.06241 & 54   &  9.9 & 76  &  11.7  & 11.13  &   9.9    & 0.55   & 317 (13)  & 18 \\   
2MASX J22073122$-$0729223     & \nodata            & 0.06331  & 60   &  8.6 & 73  &  11.6  & 11.20  & 10.3    & 1.00   & 243 (13) & 31 \\   
2MASX J23130513$-$0033477      & \nodata           &  0.11107  & 53  &   8.2 & 53  & 11.6  & 11.20  & 10.3    & 1.03   & 292 ( 8)  & 19 \\   
\enddata
\tablenotetext{a}{Redshift from SDSS DR13.}
\tablenotetext{b}{Isophotal diameter (kpc) at $\Sigma_r=25.0$ mag arcsec$^{-2}$.}
\tablenotetext{c}{Exponential disk scale length and inclination fit from SDSS images \citep{smp11}.}
\tablenotetext{d}{Dark halo mass ($M_\odot$) inside $r$, estimated from our fit to the rotation curve, assuming an NFW profile.}
\tablenotetext{e}{Mass in stars ($M_\odot$) estimated from $W1$-band luminosity, assuming $M/L$=0.6.}
\tablenotetext{f}{Gas mass ($M_\odot$) estimated from SFR and $D_{25}$, using the \cite{k98} Schmidt Law.}
\tablenotetext{g}{Star formation rate estimated from WISE 12 $\mu$m band using relation of \cite{cjd17}. }
\tablenotetext{h}{Maximum deprojected speed $v_\mathrm{max}$ (km s$^{-1}$) at radius $r$ (kpc) measured from sampled $H\alpha$ rotation curve, and standard deviation.}
\end{deluxetable*}

The specific angular momentum of galaxies increases with $M_\mathrm{stars}$, following the \cite{f83} relation $j_* \sim M^{\alpha}$ (Fig. (2b)), which is related to the BTFR via the $M_\mathrm{stars}$-radius relation. The power-law index of this scaling relation for spirals ($\alpha = 0.58 \pm 0.10$) and for ellipticals ($\alpha = 0.83 \pm 0.16$) differ from the theoretical index for dark matter halos ($\alpha = 2/3$), reflecting differences in angular momentum retention for the two types of galaxy \citep{fr13,pfd18}.  We estimate the specific angular momentum for our sample of super spirals as  $j_* = 2 v_{max} R_d$, which holds true for pure exponential disks with characteristic radius $R_d$. We take $R_d$ from \cite{smp11}, who fit the SDSS images with DeVaucouleurs bulges plus exponential disks.  We compare our sample to the spiral and elliptical samples of  \cite{fr13, fr18} and the spiral and dwarf sample of \cite{pfd18}, the latter drawn from the \cite{l16} BTFR sample. We find very high specific angular momenta which break from the Fall relation at  $\log M_\mathrm{stars}/ M_\odot \simeq 11.5 $, similar to the break we find in the Tully-Fisher relation.  
The Fall relation connects galaxy spin to halo spin, allowing us to estimate the dark halo mass of super spirals. The halo masses that are required to put super spirals on an extrapolated Fall relation range from 
$\log M_\mathrm{halo} / M_\odot= (12.0-13.6) f_\mathrm{c}/f_\mathrm{b}$. The most massive of these super spiral halos is similar to that of a typical galaxy group.

\subsection{Galaxy Mass Limit}
We suggest that the high-mass breaks in the BTFR and the Fall relation at  $\log M_\mathrm{stars} / M_\odot \simeq 11.5$ are imposed by an upper limit to the cold baryonic mass in galaxies.  Including our sample and the super spiral sample of \cite{o19}, we find a maximum baryonic mass of $ M_\mathrm{b,max} = 6.3\times10^{11} M_\odot$ ($\log M_\mathrm{b,max}/ M_\odot = 11.8 $) for super spiral OGC 0139, which is slightly more massive than the previous record-holder, ISOHDFS:[RFA2002] S27 \citep{rfa02}. The same mass upper limit may apply to elliptical and lenticular as well as spiral galaxies.  The majority of OGC giant ellipticals and super lenticulars \citep{o19} do have $M_\mathrm{stars}$ lower than the most massive super spiral (Fig. 3).  We suggest that the giant ellipticals and super lenticulars that exceed the limit (by up to a factor of 2) may be the product of major mergers.

The observed upper limit to galaxy mass in stars agrees with the theoretical prediction of the maximum halo mass where gas can cool and collapse within a dynamical time \citep{wr78}, but only if we assume that nearly all of the baryons in super spiral and giant elliptical subhalos have been incorporated into stars at the current epoch.  For an initial power-law density perturbation spectrum with amplitude $\propto M^{-1/3}$ and no metal enrichment, \cite{wr78} predict a maximum galaxy halo mass of $\log M_\mathrm{max} / M_\odot \simeq 12.7$, which is close to  the maximum enclosed dark mass in our super spiral sample ($\log M_\mathrm{d,max} / M_\odot =12.5$). For a stellar mass fraction equal to $\sim 70\%$ of cosmic baryon fraction, this corresponds to the observed maximum mass in stars of $\log M_\mathrm{stars,max} / M_\odot = 11.7$. 

The excess specific angular momentum in super spirals can be explained if it is inherited from host halos that are up to 10 times more massive than $\log M_\mathrm{max}$.
 In general, sub-halos will have lower $j_*$ values than their host halos, following the Fall relation, and when they merge they will create even lower $j_*$ elliptical galaxies.  However, a dominant central galaxy may share the specific angular momentum of its host halo if it is formed at the halo center and subsequently cools  and collapses to form a high-$j_*$ super spiral.

The continuing star formation in massive spirals appears to buck the trend of star-formation quenching in  galaxies with $L > L^*$.  This so-called "failed feedback" problem \citep {pmff19} may be resolved if  these massive spiral galaxies are 
immune to quenching. The super spirals that have survived must be robust against the various proposed quenching agents, including mergers, AGN feedback, virial shocks, and ram-pressure stripping.  If super spirals formed in dominant subhalos located at the centers of group-mass halos, they must not have suffered major mergers after the bulk of their stars were formed, and are not subject to ram-pressure stripping.  The giant gas disks of super spirals must also be immune to disruption by AGN feedback from the supermassive black holes
in their relatively small bulges.

\acknowledgements
We thank S. Michael Fall for insightful discussions about the angular momentum of galaxies and galaxy halos.
The spectroscopic observations reported in this paper were obtained with the Southern African Large Telescope (SALT) and the Hale Telescope at Palomar Observatory.
This work relied on the NASA/IPAC Extragalactic Database and the NASA/ IPAC Infrared Science Archive, which are both operated by the Jet Propulsion Laboratory, California Institute of Technology, under contract with the National Aeronautics and Space Administration.  In particular, this publication makes use of data products from the {\it Wide-field Infrared Survey Explorer}, which is a joint project of the University of California, Los Angeles, and the Jet Propulsion Laboratory/California Institute of Technology, funded by the National Aeronautics and Space Administration.

\end{document}